\documentclass[twocolumn, switch]{article} 

\usepackage{preprint}

\usepackage{amsmath, amsthm, amssymb, amsfonts}

\usepackage[numbers,square]{natbib}
\usepackage[utf8]{inputenc}	
\usepackage[T1]{fontenc}	
\usepackage{xcolor}		
\usepackage[colorlinks = true,
            linkcolor = purple,
            urlcolor  = blue,
            citecolor = cyan,
            anchorcolor = black]{hyperref}	
\usepackage{booktabs} 		
\usepackage{nicefrac}		
\usepackage{microtype}		
\usepackage{lineno}		
\usepackage{float}

\usepackage{newfloat}
\DeclareFloatingEnvironment[name={Supplementary Figure}]{suppfigure}
\usepackage{sidecap}
\sidecaptionvpos{figure}{c}

\usepackage{titlesec}
\titlespacing\section{0pt}{12pt plus 3pt minus 3pt}{1pt plus 1pt minus 1pt}
\titlespacing\subsection{0pt}{10pt plus 3pt minus 3pt}{1pt plus 1pt minus 1pt}
\titlespacing\subsubsection{0pt}{8pt plus 3pt minus 3pt}{1pt plus 1pt minus 1pt}

\usepackage{tikz,xcolor,hyperref}

\definecolor{lime}{HTML}{A6CE39}
\DeclareRobustCommand{\orcidicon}{
	\begin{tikzpicture}
	\draw[lime, fill=lime] (0,0)
	circle [radius=0.16]
	node[white] {{\fontfamily{qag}\selectfont \tiny ID}};
	\draw[white, fill=white] (-0.0625,0.095)
	circle [radius=0.007];
	\end{tikzpicture}
	\hspace{-2mm}
}
\foreach \x in {A, ..., Z}{\expandafter\xdef\csname orcid\x\endcsname{\noexpand\href{https://orcid.org/\csname orcidauthor\x\endcsname}
			{\noexpand\orcidicon}}
}

\title{Time-Varying Audio Effect Modeling by End-to-End Adversarial Training}

\usepackage{xwatermark}

\newwatermark[firstpage,color=gray!90,angle=0,scale=0.28, xpos=0in,ypos=-5in]{*correspondence: \texttt{yann.bourdin@arturia.com}}

\usepackage{authblk}

\author[1,2\thanks{\tt{yann.bourdin@arturia.com}}]{Yann Bourdin\orcidA{}}
\author[2,3]{Pierrick Legrand}
\author[1]{Fanny Roche}

\affil[1]{Arturia, F-38330 Montbonnot-Saint-Martin, France}
\affil[2]{Inria center at the University of Bordeaux, Astral Inria Team, F-33405 Talence, France}
\affil[3]{IMS, UMR CNRS 5218, ENSC, Bordeaux INP, F-33405 Talence, France}

\usepackage{amsfonts}
\usepackage{caption}
\usepackage{subcaption}
\usepackage{rotating}
\usepackage{mathtools}

\newcommand{\norm}[1]{{\lVert\, #1 \, \rVert}}
\usepackage{dblfloatfix}

\begin{document}

\twocolumn[ 
  \begin{@twocolumnfalse} 

\maketitle

\begin{abstract}
Deep learning has become a standard approach for the modeling of audio effects, yet strictly black-box modeling remains problematic for time-varying systems. Unlike time-invariant effects, training models on devices with internal modulation typically requires the recording or extraction of control signals to ensure the time-alignment required by standard loss functions. This paper introduces a Generative Adversarial Network (GAN) framework to model such effects using only input-output audio recordings, without requiring a modulation signal extraction. We propose a convolutional-recurrent architecture trained via a two-stage strategy: an initial adversarial phase allows the model to learn the distribution of the modulation behavior without strict phase constraints, followed by a supervised fine-tuning phase where a State Prediction Network (SPN) estimates the initial internal states required to synchronize the model with the target. Additionally, a new metric based on chirp-train signals is developed to quantify modulation accuracy. Experiments modeling a vintage hardware phaser demonstrate the method’s ability to capture time-varying dynamics in a fully black-box context.
\end{abstract}

\vspace{0.35cm}

  \end{@twocolumnfalse} 
]

\section{INTRODUCTION}

Audio effects are fundamental tools widely used by musicians, producers, and engineers to shape sound timbre, dynamics, and spatial characteristics.
With the widespread adoption of digital workflows, the need for the digital emulation of analog hardware has intensified, aiming to preserve the unique, often non-linear sonic characteristics of classic gear in a modern software environment.
This field, known as Virtual Analog, distinguishes three method categories, depending on their reliance on target circuit knowledge or input-output data.
White-box techniques require extensive knowledge of the target effect's topology in order to emulate it via numerical simulation or digital signal processing; they require minimal data \cite{eichasPhysicalModelingMXR2014}. In contrast, black-box techniques rely primarily on measured input-output signals \cite{ramirezGeneralpurposeDeepLearning2019, mitcheltreeModulationExtractionLFOdriven2023}. Finally, grey-box methods combine hardware knowledge and data, typically by extracting information from signal measurements to improve white-box models \cite{kiiskiTimeVariantGrayBoxModeling2016, carsonDifferentiableGreyboxModelling2023}.
Machine learning has gained significant attention for facilitating this extraction process and enabling new black-box methods capable of learning complex, non-linear mappings directly from input-output recordings. 
A large body of research in grey-box and black-box methods has successfully modeled effects with limited time-dependency, including linear effects (e.g., equalizers \cite{ramirez2018equalization}) and non-linear ones (e.g., distortions and tube amplifiers \cite{vanhataloReviewNeuralNetworkBased2022}). 
Fewer studies have tackled time-dependent effects such as compressors or specific distortions \cite{steinmetzEfficientNeuralNetworks2022, comunitaModellingBlackBoxAudio2023, bourdinTacklingLongRangeDependencies2024, bourdinEmpiricalResultsAdjusting2025}.
A less explored subject is modeling time-varying audio effects, where the internal parameters of the effect change as a function of time. Examples include chorus, flanger, and phaser effects driven by a Low-Frequency Oscillator (LFO). Several prior studies investigated the use of data-driven methods for modeling phasers. 
In \cite{kiiskiTimeVariantGrayBoxModeling2016}, the authors proposed a grey-box model by analyzing a specific test signal. In \cite{ramirezGeneralpurposeDeepLearning2019}, an autoencoder-like architecture was proposed, employing bidirectional Long Short-Term Memory (LSTM) networks in the latent space. The authors of \cite{carsonDifferentiableGreyboxModelling2023} and \cite{leeCONMODControllableNeural2024} utilize neural networks but within a frame-based spectral processing approach. Other works first extract the LFO signal, either manually \cite{wrightNeuralModelingPhaser2021} or with a neural network \cite{mitcheltreeModulationExtractionLFOdriven2023}, to use it as an additional input to an LSTM.

In the literature, ``time-varying effect'' is a frequently employed term, as in \cite{steinmetzEfficientNeuralNetworks2022, comunitaModellingBlackBoxAudio2023}, whereas their effects of interest (compression and distortion) could be considered solely time-dependent \cite{ramirezTable}. Therefore, we will rely on the following definition. Time-dependent effects are time-invariant systems, viewed as functions mapping a number of past input values to output samples.
The time dependency duration of a time-dependent effect and, as such, the number of past values affecting an output sample, varies with the effect type. It tends to be short for effects like distortions (10-100 ms) but can be long for effects such as compression or reverberation (several seconds).
Deep learning models addressing such effects can either be stateless or stateful. In stateless models, each output sample results from an operation with a finite number of past input values, corresponding to the model's receptive field \cite{steinmetzEfficientNeuralNetworks2022}. In contrast, stateful models have a theoretically infinite time dependency by employing an internal state updated recursively, often through recurrent layers such as LSTMs \cite{ramirezGeneralpurposeDeepLearning2019, leeCONMODControllableNeural2024, wrightNeuralModelingPhaser2021, mitcheltreeModulationExtractionLFOdriven2023}.
The receptive field can then be reduced, but a proper initialization of the internal state is required. It typically consists in filling the state with zeros and running the model for a number of timesteps, without gradient backpropagation, to `warm up' the states. However, long-range time dependencies may require large warm-up durations, making this inefficient. In \cite{bourdinTacklingLongRangeDependencies2024, bourdinEmpiricalResultsAdjusting2025}, a State Prediction Network (SPN) uses the past values of both the input and target output signals to predict proper initial states, replacing the use of warm-up samples with a more efficient computation. 

Time-varying effects are time-varying systems, viewed as functions of both input samples and time. They typically involve one or more of the effect's internal parameters (e.g., delay time, gain, filter cutoff frequency) modified over time by a modulation signal. When periodic, it is commonly called an LFO.
In stateful models, the internal state can naturally represent the influence of time or modulation; thus recurrent models can handle the time-varying nature of an effect.
However, the internal state now embeds information relative to the time when a signal is processed. In the supervised learning approach, input-output data are recorded at a specific time. For an LFO-based effect, it implies that each recording features a different starting LFO phase. 
However, supervised loss functions typically expect time-alignment of the compared features; therefore, the same signal processed by the same effect but with different LFO initial phases will yield a large error when compared.
Thus, without proper care, training on a dataset with varying initial LFO phases results, at best, in a model averaging the behavior of the effect over the whole LFO period.
Other works attempting end-to-end modeling of LFO-based effects either limit their dataset and adapt the training \cite{carsonDifferentiableGreyboxModelling2023, leeCONMODControllableNeural2024}, train over a dataset with a fixed LFO signal \cite{ramirezGeneralpurposeDeepLearning2019}, or pre-train an LFO extraction network \cite{mitcheltreeModulationExtractionLFOdriven2023}.
In our work, we extend the state prediction approach -- originally designed for time-dependent effects -- to time-varying ones, and propose an adversarial approach enabling training on large datasets with fewer constraints and without depending on LFO extraction, but at the cost of modulation being more difficult to interpret and control. Additionally, although not yet supported by empirical evidence, our method is designed with the intent to apply to any time-varying effect, not necessarily LFO-driven.

The paper is structured as follows. Section \ref{generative} reviews Generative Adversarial Networks. Section \ref{method} details the proposed method, describing the generator and discriminator architectures, the SPN, and the training strategy and novel modulation metric. Section \ref{experimental} outlines the experimental setup, including the target effect and dataset creation. Section \ref{experiments} presents results regarding state initialization, mode seeking, and the two-phase training efficacy. Section \ref{conclusion} concludes.

\section{GENERATIVE ADVERSARIAL NETWORKS}
\label{generative}

Generative Adversarial Networks (GAN) are generative models obtained through the concurrent optimization of two models called the generator and discriminator (or critic).
The discriminator is a binary classifier distinguishing between ``true" samples from the dataset and ``fake" samples from the generator.
The generator aims to generate samples resembling the target data, fooling the discriminator.
The GAN framework can be formulated as a two-player minimax game \cite{goodfellow2014generative}:  
\begin{equation}
    \operatorname*{min}_{G}\operatorname*{max}_{D} \mathbb{E}_{x\sim p_d(x)}[\log D(x)]+\mathbb{E}_{z\sim p_{z}(z)}[\log(1-D(G(z)))]
\end{equation}
\noindent where $p_d$ represents the distribution over the training data, and $p_z$ the distribution over the generator's input variables.
In practice, this optimization is performed separately by alternating minimization steps between the generator and the discriminator:
\begin{align}
    \operatorname*{min}_{D} \; & - \mathbb{E}_{x\sim p_d(x)}[\log D(x)] - \mathbb{E}_{z\sim p_{z}(z)}[\log(1-D(G(z)))] \nonumber \\
    \operatorname*{min}_{G} \; & \mathbb{E}_{z\sim p_{z}(z)}[\log(1-D(G(z)))]\: .
\end{align}
However, following this original formulation, numerous improvements have been proposed, notably regarding the nature of the adversarial losses, incorporating regularization terms \cite{arjovsky2017towards, thanh2020catastrophic}, as well as normalization methods \cite{miyato2018spectral}. 
Though GANs were initially proposed for image domain tasks, they have seen extensive success in audio tasks such as speech synthesis and neural vocoding \cite{kumarMelGANGenerativeAdversarial2019, kongHiFiGANGenerativeAdversarial2020}, as well as neural compression \cite{defossezHighFidelityNeural2022}. 
Key architectural advancements in GANs for audio include specialized discriminators such as the Multi-Scale Discriminator (MSD) \cite{kumarMelGANGenerativeAdversarial2019} and multi-resolution STFT-based discriminators \cite{defossezHighFidelityNeural2022}, handling audio at different resolutions, as well as the Multi-Period Discriminator (MPD) \cite{kongHiFiGANGenerativeAdversarial2020} to better capture the periodic structures inherent in audio signals.
With such discriminators, GAN models are able to learn data-driven loss functions which can improve the details and perceptual quality of the generated audio \cite{defossezHighFidelityNeural2022, chenImprovingUnsupervisedCleantoRendered2024}.
In audio effect modeling, GANs have been applied to guitar amplifier modeling \cite{chenImprovingUnsupervisedCleantoRendered2024, wrightAdversarialGuitarAmplifier2023}, especially when paired input-output data is unavailable.

In the context of this study, input-output pairs are available. Our dataset contains tuples ($x$, $y$, $\varphi$) of input $x$ and output $y$ audio slices and the corresponding vector of user controls $\varphi$. The generator processes batches of tuples $(x, \varphi, h^0)$, where $h^0$ corresponds to initial states, to produce output signals $\hat{y}$. The discriminator evaluates ``true'' tuples $D(x, \varphi, y)$ and generated tuples $D(x, \varphi, \hat{y})$.

\section{METHOD}
\label{method}

\begin{figure}
    \centering
    \begin{subfigure}[b]{0.49\textwidth}
     \centering
     \includegraphics[width=0.75\textwidth]{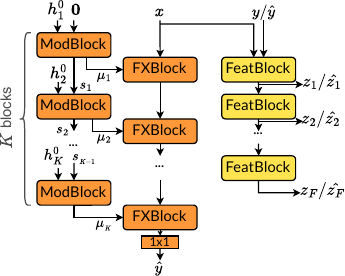}
     \caption{Modulation, audio processing, and feature extraction paths.}
     \label{fig:sptvmod1}
    \end{subfigure}
    \\
    \begin{subfigure}[b]{0.49\textwidth}
     \centering
     \includegraphics[width=\textwidth]{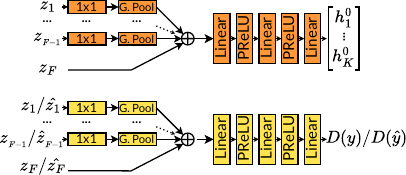}
     \caption{State Prediction Network (in orange) and the discriminator's head (in yellow), processing the features $z_j$ and $\widehat{z_j}$.}
     \label{fig:sptvmod2}
    \end{subfigure}
    \caption{Block diagram of the architecture of SPTVMod. The generator has an orange background, and the discriminator a yellow one.}
    \label{fig:sptvmod}
\end{figure}

\subsection{Generator}
\label{generator}

\begin{figure}
    \centering
    \includegraphics[width=0.48\textwidth]{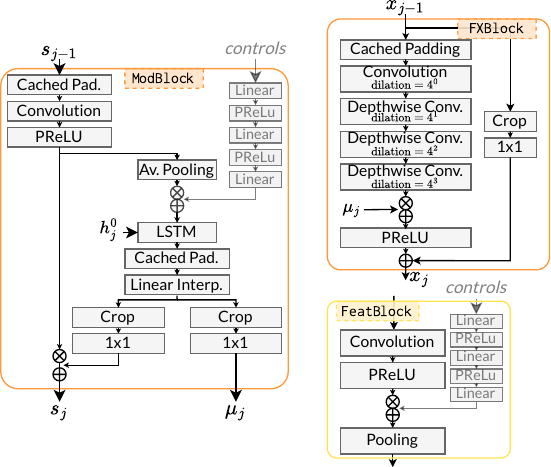}
    \caption{Composition of the processing blocks: ModBlock (left), FXBlock (upper right) and FeatBlock (lower right).}
    \label{fig:blocks}
\end{figure}

Although our framework is compatible with any supervised audio effect modeling architecture, we employ the Series-Parallel Time-Varying Modulation (SPTVMod) network, depicted in Figure \ref{fig:sptvmod}. Fig. \ref{fig:sptvmod1} shows its main processing blocks, and is completed by Fig. \ref{fig:sptvmod2} detailing the discriminator and the SPN, to be introduced in Sections \ref{discriminator} and \ref{state}. Derived from the SPTMod architecture proposed in \cite{bourdinTacklingLongRangeDependencies2024, bourdinEmpiricalResultsAdjusting2025}, it features a modulation path and an audio path, the former computing modulation signals applied to the latter. Compared to SPTMod, SPTVMod introduces processing blocks into the audio path, called FXBlocks; and the modulation path's first block no longer receives the audio input but a zero-filled signal.

In this work, conditioning by user controls was not investigated, but we acknowledge that conditioning using Feature Wise Linear Modulation (FiLM) \cite{steinmetzEfficientNeuralNetworks2022, perezFiLMVisualReasoning2017} can be done in a similar way as in SPTMod and is illustrated in Fig. \ref{fig:blocks} with greyed-out components. While the architecture is thus suited for full modeling with user controls, we limit this study to `snapshot' datasets, containing a single set of user controls, resulting in FiLM being bypassed.

The modulation blocks, ModBlocks, detailed in Fig. \ref{fig:blocks} (left), are the same as in \cite{bourdinEmpiricalResultsAdjusting2025}. Each block begins with a convolutional layer and PReLU activation, feeding a recurrent branch that undergoes FiLM conditioning (if active) and contains an LSTM layer. To handle timesteps at a lower sample rate, the LSTM is surrounded by Average Pooling (downsampling) and Linear Interpolation (upsampling) layers. The branch terminates with two parallel $1\times1$ layers (convolutional layers with a kernel size of one). One generates the modulation tensors $\mu_j$ for the audio path, and the other produces a residual added to the block's main branch to form $s_j$, the input for the next ModBlock.

The FXBlocks, detailed in Fig. \ref{fig:blocks} (upper right), consist of a residual connection around a sequence of convolutional layers, whose output is modulated by the tensor $\mu_j$ received from the adjacent ModBlock via a FiLM operation, followed by a PReLU activation. We apply an exponentially increasing dilation factor to these convolutions. While typically used to expand the receptive field \cite{steinmetzEfficientNeuralNetworks2022}, our motivation lies in signal processing: here dilation extends the filters' response length, improving spectral accuracy and low-frequency processing. To keep the parameter count and computational cost low, we employ depthwise convolutions for all layers following the first, i.e. each input channel undergoes a single convolution.

The LSTM layers are initialized with the states $h^0_j$. In our GAN framework, the $h^0_j$ introduce the necessary stochasticity for generation; they are either sampled from a probability distribution or predicted by an SPN, to be introduced in Section \ref{state}. 

The cached padding and cropping layers account for the reduction in temporal length caused by operations like convolutions and pooling. To avoid artefacts associated with zero-padding, we compute the precise input length needed so that the model outputs the target length, as done in \cite{bourdinTacklingLongRangeDependencies2024, bourdinEmpiricalResultsAdjusting2025}. However, the architecture retains a cached padding mechanism \cite{caillonStreamableNeuralAudio2022} to facilitate streaming inference and potentially using Truncated Backpropagation Through Time \cite{bourdinEmpiricalResultsAdjusting2025}.

\subsection{Discriminator}
\label{discriminator}

Prior to the adoption of the adversarial framework, we investigated the use of an SPN \cite{bourdinTacklingLongRangeDependencies2024, bourdinEmpiricalResultsAdjusting2025} for a fully supervised setup. Given its structure as a convolutional classifier, the architecture is naturally suited to serve as a discriminator's backbone. We adapt this idea by constructing our discriminator from a series of feature extraction blocks, named FeatBlocks and depicted in Fig. \ref{fig:blocks} (lower right). It consists of a convolution, PReLU activation, FiLM conditioning (if active) and average pooling, extracting hierarchical features $z_j$ (for real data) or $\widehat{z_j}$ (for generated data).
The pooling layers progressively reduce the time length of the features. Subsequently, as depicted in Fig. \ref{fig:sptvmod2}, each feature $z_j$ (or $\widehat{z_j}$) is processed by a $1\times1$ convolution and a global average pooling layer (averaging over the whole duration). These processed tensors are summed and finally passed through a fully connected network with PReLU activations to yield the discriminator's scalar output.

\subsection{State Prediction}
\label{state}

We use an SPN to initialize the LSTM states $h^0_j$, adapted here for time-varying effects, and detailed in Fig. \ref{fig:sptvmod2}. Unlike previous work where the SPN processed values preceding the target \cite{bourdinTacklingLongRangeDependencies2024}, our SPN analyzes the current processing window $(x, y, \varphi)$, comprising both input and output signals.
We leverage the discriminator's architecture for feature extraction. Since the discriminator already processes $(x, y, \varphi)$ tuples, we re-use its learned features to drive state prediction. Our preliminary results indicated this was more efficient than instantiating a duplicate architecture. Furthermore, this acts as a form of transfer learning, facilitating the SPN's training.

Therefore, the SPN comprises $1\times1$ layers and global average pooling layers (similar to the discriminator) that process the discriminator's features, followed by a neural network determining the initial states $h^0_j$. The parameters of these layers are updated by the generator's optimizer. When using an SPN in this setup, the discriminator's parameters should be frozen to ensure that its features remain stable; otherwise, our preliminary experiments showed that the SPN would fail to learn properly.

Note that since the SPN requires the ground truth, there is a risk of exposure bias \cite{peussaExposureBiasState2021}, which was mitigated in previous work \cite{bourdinEmpiricalResultsAdjusting2025}. Also, the SPN is only intended to facilitate training and is not necessary at inference, except if precise initial states are needed; in this study, to obtain comparable results, we use the SPN during evaluation, see Section \ref{training_and}. 

\subsection{Training Method}
\label{Training}

\subsubsection{Loss Functions}
\label{loss}

The generator and the discriminator are trained using the hinge loss objective, given its success in prior work \cite{kumarMelGANGenerativeAdversarial2019, defossezHighFidelityNeural2022, chenImprovingUnsupervisedCleantoRendered2024, wrightAdversarialGuitarAmplifier2023}. The generator minimizes:
\begin{equation}
    \mathcal{L}^G_{\mathrm{adv}} = \operatorname{max} (0, 1 - D(x, \varphi, \hat{y}))
\end{equation}
while the discriminator minimizes:
\begin{equation}
    \mathcal{L}^D_{\mathrm{adv}} = \operatorname{max} \left (0, 1 - D(x, \varphi, y)\right )
     + \operatorname{max} \left (0, 1 + D(x, \varphi, \hat{y})\right )\: .
\end{equation}
Previous studies demonstrate that including supervised losses, over $y$ and $\hat{y}$, and feature-matching losses, over the true and fake features of the discriminator ($z_j$ and $\widehat{z_j}$), can facilitate the training of GANs \cite{kumarMelGANGenerativeAdversarial2019, defossezHighFidelityNeural2022}.
We include a Multi-Resolution Short-Term Fourier Transform (MR-STFT) loss to the generator, to be referred to as spectral loss. It is a mean of multiple $\mathcal{L}^G_{\textrm{STFT}}$ losses, defined below, computed using different STFT window sizes (512, 1024 and 2048 samples):
\begin{equation}
    \mathcal{L}^G_{\textrm{STFT}} = \frac{\norm{|Y| - |\hat{Y}|}_F}{\norm{|Y|}_F} + \frac{1}{L} \norm{\log |Y| - \log |\hat{Y}|}_1
\end{equation}
where $L$ is the STFT window size, $Y$ and $\hat{Y}$ are the STFTs of $y$ and $\hat{y}$, and $\norm{.}_F$ and $\norm{.}_1$ are the Frobenius and $l_1$ norms.
This loss function is widely used in audio processing literature for its discrimination of time-frequency features, resembling human perception, and is robust to phase mismatches between signals.
However, using a feature-matching loss was not considered in this study as our preliminary experiments indicated that it negatively impacted the training in our context.

In order to enhance the stability of the adversarial training, since the gradient of $\mathcal{L}^G_{\mathrm{adv}}$ can vary over several orders of magnitude, we employ the adaptive loss balancing technique proposed in \cite{defossezHighFidelityNeural2022} which facilitates the weighting of several loss terms. It consists in rescaling the gradients of each term of the generator's loss at every iteration indexed by $t$, according to:
\begin{equation}
    {g_i^t}^\prime = \frac{\lambda_i}{\sum_j \lambda_j} \cdot \frac{g_i^t}{<\norm{g_i}_2>^t}
\end{equation}
where $g_i^\prime$ is the corrected gradient of the $i$-th generator loss, associated to a weight $\lambda_i$ and the original gradient $g_i$, and $<.>$ is an exponential moving average with a rate $\alpha = 0.99$, that is:
\begin{equation}
    <\norm{g_i}_2>^t = (1 - \alpha) \norm{g_i}_2^t + \alpha \norm{g_i}_2^{t-1} \: .
\end{equation}

\subsubsection{Training Strategy}
\label{training}

\begin{figure}
    \centering
    \includegraphics[width=\linewidth]{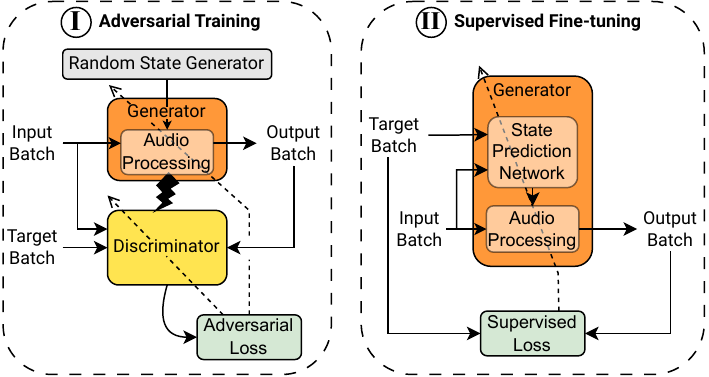}
    \caption{Illustration of our two-phase training strategy. The dotted arrows represent the backpropagation of the loss gradients.}
    \label{fig:training}
\end{figure}

Preliminary experiments revealed that a purely \mbox{adversarial} training faced significant stability issues. We observed recurring imbalances between the generator and the discriminator, alongside signs of catastrophic forgetting \cite{thanh2020catastrophic}: despite initially capturing the modulation characteristics, the generator would occasionally undergo sudden performance degradation, losing the learned modulation behavior and requiring many iterations to recover. Furthermore, the SPN failed to converge in the adversarial context, possibly due to the instable learning of modulation patterns. With LFO-driven effects, the learned LFO frequency keeps varying during training, making the prediction of its initial phase difficult.
To mitigate these issues, we propose a two-phase training strategy, illustrated in Fig. \ref{fig:training}.

\paragraph*{Phase I: Adversarial Training.} The first phase focuses on training the model using primarily adversarial losses. State prediction is disabled; instead, the internal states $h^0_j$ are initialized stochastically. This forces the model to learn the time-varying effect's general behavior without relying on a specific LFO phase. We found that retaining a spectral loss with a very low weighting was beneficial for stability.

\paragraph*{Phase II: Supervised Fine-tuning.} The second phase involves removing the adversarial objective, rendering discrimination unnecessary, and optimizing the model using the spectral loss combined with the SPN. Here, the SPN learns to synchronize the model's internal modulation with the target data. This approach stabilizes the training by providing a perceptually motivated objective, subsequently serving as the stopping criterion and evaluation metric.

\subsubsection{Mode Seeking}
\label{mode}

Our framework is that of Conditional GANs \cite{mirza2014conditional}, where the generation process is conditioned on the input audio signal and potentially user control parameters. Despite their success in various applications, conditional GANs are notably susceptible to mode collapse, a phenomenon where the generator yields a limited subset of outputs, failing to capture the full diversity of the target data. This issue is common in conditional settings: because the high-dimensional conditioning signals provide strong structural priors, the generator tends to disregard the lower-dimensional stochastic input vector intended to induce variation, here embodied by the initial states $h^0_j$. 
In the specific context of our work, this phenomenon manifests when the generator's internal states fail to influence the time-varying behavior of the model, yielding a fixed initial modulation phase regardless of the stochastic input. However, it seems essential that the modulation remains sensitive to the initial states; otherwise the SPN would be unable to synchronize the modulation phase with that of the training data during the subsequent fine-tuning stage.

To mitigate mode collapse, various techniques have been proposed, including Wasserstein-GAN regularization \cite{arjovsky2017towards, thanh2020catastrophic} and spectral normalization \cite{miyato2018spectral} methods. Despite our efforts to utilize these techniques, they introduced difficulties in training without mitigating mode collapse. For our application, we adopt a strategy based on the mode seeking method introduced in \cite{maoModeSeekingGenerative2019}, consisting in a regularization term added to the generator's objective to explicitly penalize mode collapse. This term maximizes the ratio of the distance between generated samples to the distance between their corresponding latent codes.
We adapt this approach by minimizing the inverse of the MR-STFT distance between signals generated from the same input but with different initial states. Rather than normalizing by the distance in the latent space, we employ a sensitivity parameter $\varepsilon$. The resulting mode seeking loss is defined as:
\begin{equation} 
\label{eq:ms}
\mathcal{L}^G_{\mathrm{ms}} = \frac{\varepsilon}{\varepsilon + \mathcal{L}_{\mathrm{MRSTFT}}\left (G(x_:, h^0_:), G(x_:, {h^0_:}^\prime)\right )}
\end{equation}
\noindent where $x_:$ represents a slice of the input signal batch, and $h^0_:$ and ${h^0_:}^\prime$ denote two distinct samplings of the corresponding initial states. During the initial adversarial phase, both are sampled randomly. However, in the presence of the SPN (during the fine-tuning phase), ${h^0_:}^\prime$ is obtained by permuting the batch of initial states $h^0$ predicted by the SPN.

This loss function is particularly well-suited for LFO-based effects, as signals processed with different initial LFO phases inherently exhibit large spectral errors, resulting in a low mode seeking loss. Consequently, this objective not only prevents mode collapse but also encourages the SPN to diversify its predicted initial states.

\subsection{Modulation Metric}
\label{modulation}

\begin{figure*}
    \centering
    \begin{subfigure}[b]{0.325\textwidth}
        \includegraphics[width=\textwidth]{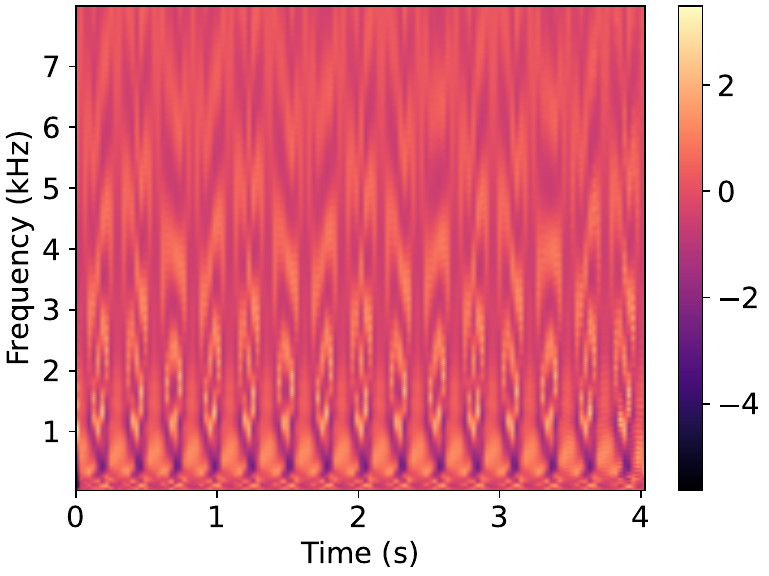}
        \caption{Chirp-aligned normalized spectrogram}
        \label{fig:jaes_mod_illus_1}
    \end{subfigure}
    \begin{subfigure}[b]{0.325\textwidth}
        \includegraphics[width=\textwidth]{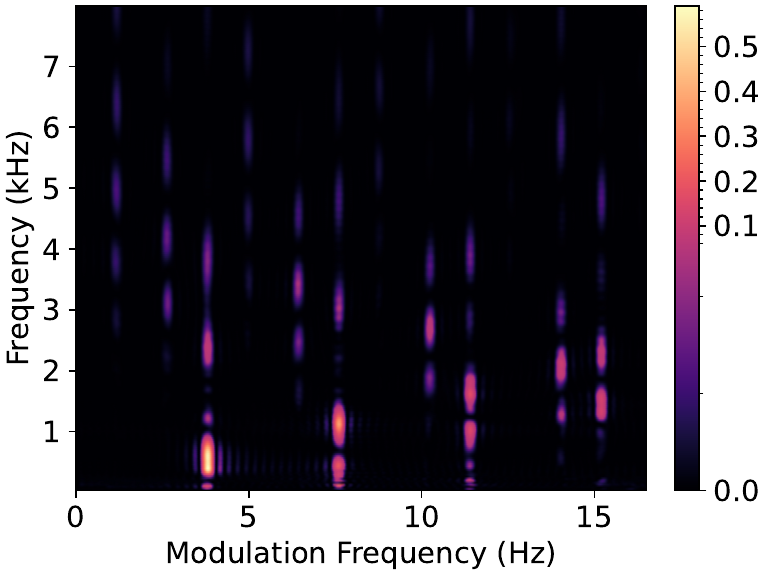}
        \caption{Frequency-frequency representation}
        \label{fig:jaes_mod_illus_2}
    \end{subfigure}
    \begin{subfigure}[b]{0.325\textwidth}
        \includegraphics[width=\textwidth]{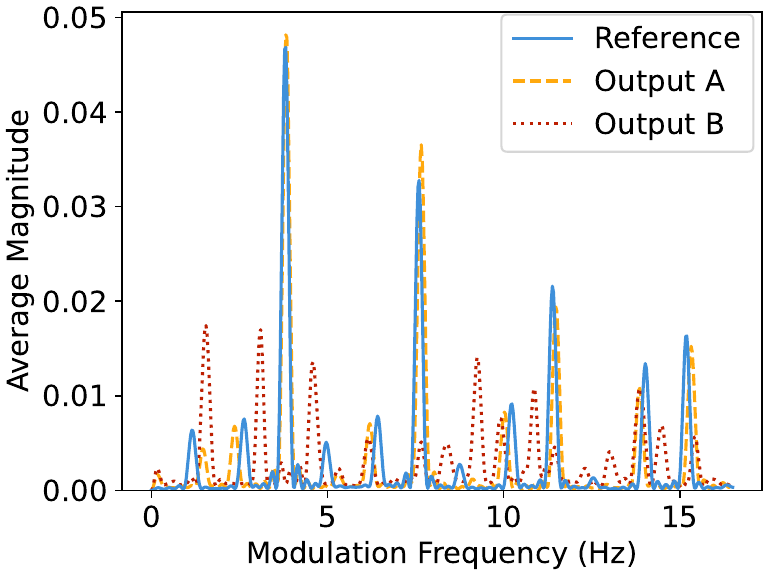}
        \caption{Modulation spectrum}
        \label{fig:jaes_mod_illus_3}
    \end{subfigure}
    \caption{Steps producing a modulation spectrum. The reference, shown in (a--c), and outputs A and B correspond respectively to the spectrograms shown further in Fig. \ref{fig:jaes_spec_long_08}, \ref{fig:jaes_spec_long_09} and \ref{fig:jaes_spec_long_11}.}
    \label{jaes_mod_illus}
\end{figure*}

In the case of an LFO-driven effect, the adversarial training enables the model to learn the LFO shape without the need to synchronize its LFO phase with that of the target. However, this limits the relevance of the MR-STFT metric, which cannot be used to determine when the adversarial training should stop. As to be discussed in the results (Section \ref{experiments}), a low MR-STFT value in this phase may be linked to an absence of modulation in the model's outputs, rather than a properly learned time-variation. Moreover, while the adversarial loss values can indicate whether there is a balance between the generator and the discriminator, they do not constitute an objective metric, although the models tend to have interesting behaviors when the generator and discriminator are close to a Nash equilibrium.

\begin{figure}
    \centering
    \includegraphics[width=0.48\textwidth]{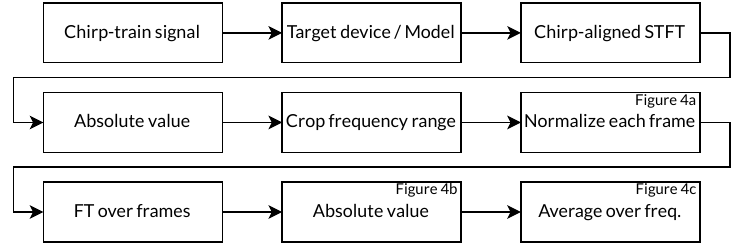}
    \caption{Process for extracting a modulation spectrum.}
    \label{fig:modspec}
\end{figure}

We propose a metric designed to be sensitive to the presence of a periodic modulation in a time-varying effect. It relies on a chirp-train signal, a test signal commonly used to analyze time-varying effects, which consists in the repetition of a sine wave with a logarithmic frequency sweep, generated by the formula:
\begin{equation}
\label{eq:chirp}
    x[n] = \sin \left (2\pi f_0 \frac{k^{\frac{n}{f_s}} - 1}{\log k} \right ) \quad \text{where} \; k = \left (\frac{f_1}{f_0} \right )^{\frac{f_s}{N}}
\end{equation}
\noindent where $N$ is the number of signal samples, $f_s$ the sampling rate, $f_0$ the starting frequency, and $f_1$ the final frequency.
This test signal enables us to sample the time-varying transfer function of the target effect, at a rate equal to the inverse of the chirp duration. 
For an effect such as an LFO-modulated phaser, the time-frequency representation of its output should exhibit periodic patterns.
We can compute a spectrum over each chirp to produce a chirp-aligned spectrogram (Fig. \ref{fig:jaes_mod_illus_1}). Each frequency in the amplitude spectrogram is a signal itself which, if affected by the LFO, should evolve at the LFO rate. Therefore, we apply a Fourier transform to each frequency in the amplitude spectrogram to obtain a frequency-frequency representation (Fig. \ref{fig:jaes_mod_illus_2}). In the presence of an LFO, we expect the amplitudes of many audio frequencies to have a large peak at the same modulation frequency in their modulation spectrum. On the frequency-frequency representation, this corresponds to columns of intensity localized at the LFO frequency and its harmonics. Finally, we sum over the audio frequency dimension to obtain a modulation spectrum, as shown in Fig. \ref{fig:jaes_mod_illus_3}, exhibiting a large peak centered around the LFO frequency. Fig. \ref{fig:modspec} summarizes our procedure.
We formulate our metric through the following criteria based on the modulation spectrums $\mathcal{M}(y)$ and $\mathcal{M}(\hat{y})$:

\begin{equation}
    \mathcal{L}_{\mathrm{mod,p}}=\frac{\left| \sum_m \mathcal{M}(y)[m] - \sum_m \mathcal{M}(\hat{y})[m] \right|}{\sum_m \mathcal{M}(y)[m]}
\end{equation}

\begin{equation}
    \mathcal{L}_{\mathrm{mod,w}} = \sum_i \left | F_{\mathcal{M}(y)}^{-1}(r_i) - F_{\mathcal{M}(\hat{y})}^{-1}(r_i) \right | (r_i - r_{i-1})
\end{equation}
\noindent where $F_{\mathcal{M}(y)}^{-1}$ is the inverse of the cumulative distribution function of $\mathcal{M}(y)$, $r$ is the ordered set of quantiles of the modulation spectrums.
The $\mathcal{L}_{\mathrm{mod,p}}$ term evaluates the presence of modulation by comparing the total energies in the modulation spectrums; while $\mathcal{L}_{\mathrm{mod,w}}$ aims to evaluate the horizontal shift between the modulation spectrums, and is an approximation to the one-dimensional 1-Wasserstein distance. This function is derived from Optimal Transport in \cite{torresUnsupervisedHarmonicParameter2024}, and we use the implementation shared by the authors.
In order to combine both criteria into a single metric, we first generated a number of outputs through several models at multiple epochs. We then observed their spectrograms and modulation spectrums, and ordered each example according to our preferences. We finally combined the terms by adjusting a coefficient, so that the ordering of the metric values would preserve our own ordering by preference. This resulted in the following modulation metric:
\begin{equation}
    \mathcal{L}_{\mathrm{mod}} = \mathcal{L}_{\mathrm{mod,p}} + 1.5\mathcal{L}_{\mathrm{mod,w}}\: .
\end{equation}

We acknowledge that the empirical nature of this weighting is a limitation of our study, compromising cross-study reproducibility, and that determining a relationship between this metric and perception would be interesting. Here we use this metric mainly to monitor the appearance of modulation in our models' outputs, to be explained in Section \ref{training_and}.
Though this metric is differentiable, we cannot use it as a training objective because of its reliance on the specific chirp-train signal, and that it requires signal windows larger than several LFO periods. This metric could however be adapted for training in grey-box approaches.

\section{EXPERIMENTAL SETUP}
\label{experimental}

\subsection{Target Effect}

We focus on the phaser effect of the Ensoniq DP/4 
digital multi-effect unit \cite{johnson1992parallel}. It is controlled by the LFO rate, width, and center frequency, as well as the Notch filter depth and feedback amount. In this article we only investigate snapshot modeling, i.e. modeling a single user control configuration per model. We fixed the LFO width to 50\%, LFO center to 50\%, filter depth to 100\%, and feedback amount to 0. Our experiments are done over two datasets using an LFO rate set to either 50\% (corresponding to a $1.3$~s period) or 80\% ($0.3$~s), respectively referred to as Slow-LFO and Fast-LFO datasets for the rest of the paper.

\subsection{Dataset Recording}

We prepared two input audio files, reserved for training and for validation. Each file contains the 4-second-long chirp-train signal,

described by (\ref{eq:chirp}) and necessary for the modulation metric, and a concatenation of 4-second music excerpts taken from the Free Music Archive dataset \cite{defferrardFMADatasetMusic2017}. The chirps have a frequency decreasing from 22.05 kHz to 20 Hz and a duration of 1/33 of a second, and are concatenated. Each chirp is multiplied by a ramp going from one to zero at the end of the chirp.
The training audio is 620 seconds long, whereas the validation audio is 144 seconds long. A gain envelope linearly increasing from -20 to 0 dB is applied to the chirp-train signal, as well as on the whole musical section. 
Both datasets' input-output pairs were recorded using the same input audio files, at a sample rate of 44.1 kHz. Because of its internal sample rate, the hardware exhibits an aliasing frequency of 17.8 kHz. Consequently, independently of the proposed method, the input files were pre-filtered using a low-pass filter before reaching the device. Similarly, the output recordings were filtered by a Finite Impulse Response (FIR) filter with 1024 taps, designed via the Remez algorithm \cite{remez} with a 17.8 kHz cutoff and a 2\% transition band to achieve -100 dB stopband attenuation. This identical filter was also added at the end of SPTVMod. This type of filter was chosen for its efficiency on Graphical Processing Units (GPU).

\subsection{Training and Evaluation Setup}
\label{training_and}

SPTVMod is implemented and trained using the PyTorch deep learning library. Its source code and complementary material to this article are available on the companion website\footnote{\url{https://ybourdin.github.io/sptvmod}}. 
Training batches are formed from windows of size 32768 of the training data, with a hop length of 4096. We use a batch size of 16 and the Adam optimizer \cite{kingmaAdamMethodStochastic2017} for both the generator and discriminator, with respective learning rates of $5\times 10^{-4}$ and $1\times 10^{-3}$. The validation loss is evaluated every 2085 steps (5 epochs) at which moment a `checkpoint', a file capturing the state of the model's parameters and optimizers, is saved.

SPTVMod's audio processing path comprises three blocks containing sequences of four convolutional layers having 16 channels and a kernel size of 16. We chose a dilation base of four, i.e. the dilation factor of the $k$-th convolution in a block is equal to $4^{k-1}$. The resulting response length is 1357 samples (31 ms at a sample rate of 44.1 kHz). In the modulation path, the ModBlocks' convolutions have 16 channels and a kernel size of 16, without dilation. The size and stride of the pooling window are 64, and the hidden size of the LSTM is 32. 

The discriminator has six FeatBlocks with a pooling window size of 4, and its convolutions use 24 channels. The three first convolutions have a kernel size of 8, the fourth has 12, and the last two have 16. These values were chosen with the aim of reducing information as efficiently as possible, resulting in a final length of 3.
The fully-connected feature-processing networks of both the discriminator and the SPN contain two layers of 24 neurons. 

\begin{figure*}
    \centering
    \includegraphics[width=0.25\textwidth]{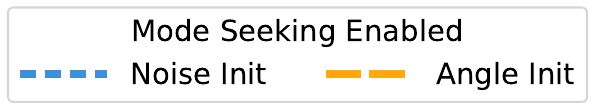}
    \hspace{4em}
    \includegraphics[width=0.25\textwidth]{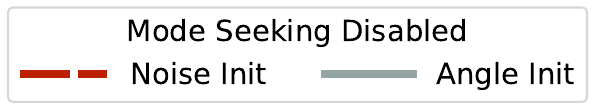}
    \\
    \vspace{-0.25em}
    \begin{subfigure}[b]{0.31\textwidth}
        \centering
        \includegraphics[width=\textwidth]{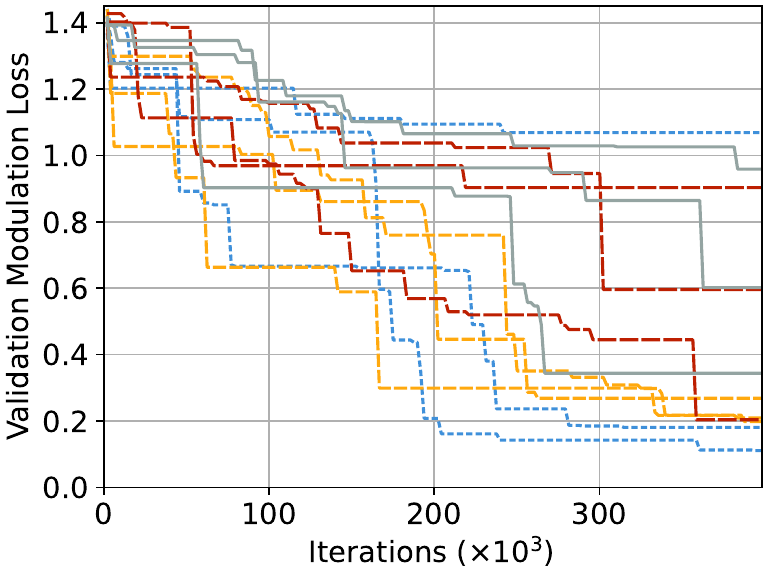}
        \caption{Running minimum of the modulation metric over the Slow-LFO dataset}
        \label{fig:g11_loss_rate50}
    \end{subfigure}
    \hfill
    \begin{subfigure}[b]{0.31\textwidth}
        \centering
        \includegraphics[width=\textwidth]{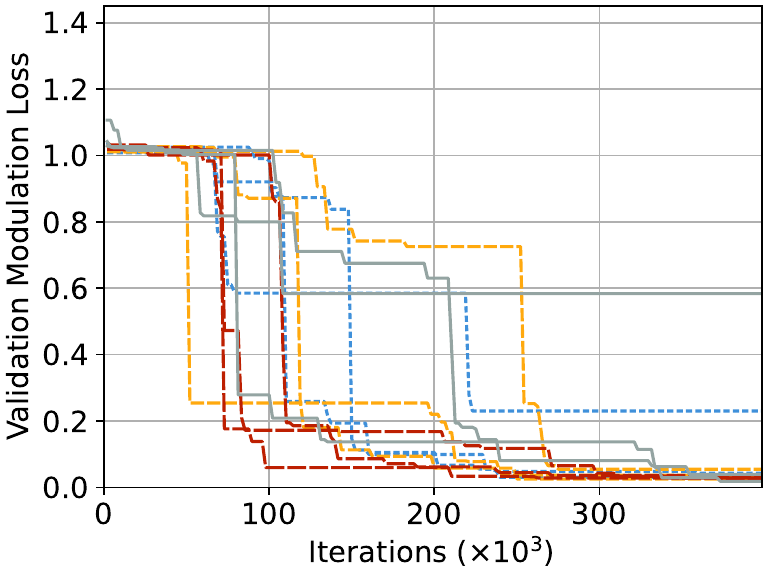}
        \caption{Running minimum of the modulation metric over the Fast-LFO dataset}
        \label{fig:g11_loss_rate80}
    \end{subfigure}
    \hfill
    \begin{subfigure}[b]{0.35\textwidth}
        \centering
        \includegraphics[width=\textwidth]{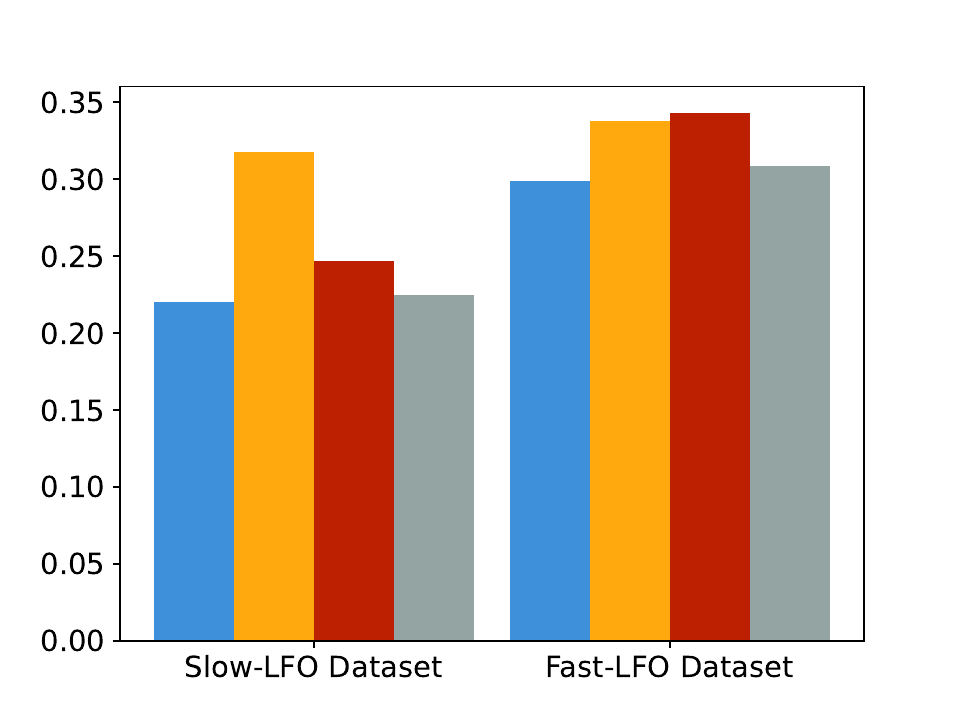}
        \caption{Mean of the difference between the modulation metric values and their running minimum.}
        \label{fig:g11_loss_diff}
    \end{subfigure}
    \caption{Evolution of the modulation metric during the adversarial phase. The presence of mode seeking, as well as the noise and angle state initialization methods are compared.}
    \label{fig:g11_loss}
\end{figure*}

The frequency range evaluated by the MR-STFT loss was limited to the aliasing limit of the device (17.8 kHz).
The MR-STFT metric is evaluated in a Windowed Target (WT) fashion: like during training, the validation dataset is split into windows of length 32768 (743 ms), the SPN processes all windows and the MR-STFT loss is computed over every window then averaged. Streamed Target (ST) differs from WT in that the audio files are processed in consecutive windows without state reset, with the SPN initializing states for the first window only. Though ST represents our final goal when modeling audio effects, mirroring inference, and that previous work addresses the WT-ST discrepancy \cite{bourdinEmpiricalResultsAdjusting2025}, we only consider WT in this study as a first step towards modeling time-varying effects.

\section{EXPERIMENTS AND RESULTS}
\label{experiments}

\subsection{Experiment 1: State Initialization Methods and Mode Seeking}
\label{experiment1}

Since the SPN is absent in the adversarial phase, the initial states must be generated stochastically. We propose and compare two approaches in this first experiment. In both cases, a stochastically generated batched vector $u$ is processed by a trainable linear layer to produce the states $h^0_j$. In the first method, which we call \emph{normal initialization}, the elements of $u$ are sampled from a standard normal distribution. In the second method, which we call \emph{angle initialization}, a batch of angles $\theta$ is sampled from a uniform distribution $\mathcal{U}[0, 2\pi]$, and $u$ is formed by $u = \left[ \cos \theta \; \sin \theta  \right]^{\mathrm{T}}$.

Our first experiment assesses the impact of the normal and angle initialization methods, as well as the absence of mode seeking or its presence. When present, we set its hyperparameters $\lambda_\mathrm{ms}$ (regularization weight) and $\varepsilon$ (appearing in Equation \ref{eq:ms}) to $\varepsilon=0.01$ and $\lambda_\mathrm{ms}=0.01$. Specifically, as depicted in Fig. \ref{fig:g11_loss}, we observe their impact on the adversarial training phase in terms of the modulation metric. The four combinations are trained three times, for each dataset, using different random seeds. 
Fig. \ref{fig:g11_loss_rate50} and \ref{fig:g11_loss_rate80} display, for both datasets, the running minimum of the modulation metric across iterations (an iteration is considered to be a step of the generator's optimizer).
On the Slow-LFO dataset (Fig. \ref{fig:g11_loss_rate50}) with mode seeking disabled, convergence was more difficult for both initialization methods. When mode seeking was enabled, over this dataset, convergence seemed faster and the two methods performed similarly, but a model using noise initialization failed to converge. On the Fast-LFO dataset however (Fig. \ref{fig:g11_loss_rate80}), all models converged without significant differences between the categories.

We noticed that during training the modulation metric tends to significantly increase after it reaches a new minimum, indicating the instability of the adversarial training. Given that this behavior is obscured by the running minimum operation, Fig. \ref{fig:g11_loss_diff} presents the mean difference between the modulation metric and its running minimum. The gaps between the categories do not seem significant, suggesting that mode seeking neither particularly improves training stability nor does the state initialization method.

\subsection{Experiment 2: Mode Seeking and Fine-Tuning}
\label{experiment2}

Our second and larger experiment applies the two-step training process detailed in Section \ref{training}, and assesses the impact of the presence of mode seeking as well as two variations in the fine-tuning phase, to be detailed in Section \ref{finetuning}.
For mode seeking, we consider a grid of its hyperparameters with $\lambda_\mathrm{ms} \in \{ 0.01, 0.1, 1 \}$ and $\varepsilon \in \{ 0.001, 0.01, 0.1 \}$. For each $(\lambda_\mathrm{ms}, \varepsilon)$ pair, three instances of the models were trained independently using different random seeds, for each dataset. For a fairer assessment, the case where there is no mode seeking ($\lambda_\mathrm{ms} = 0$) is evaluated with nine instances per dataset.

\subsubsection{Adversarial Phase}

\def\specdesc{The spectrograms use boxcar chirp-aligned windows, i.e. the window and hop sizes are equal to the chirp length (1336 samples); the frequency axis ranges from 0.04 to 17 kHz in log-scale.}

\begin{figure}
    \centering
    \includegraphics[width=0.4\textwidth]{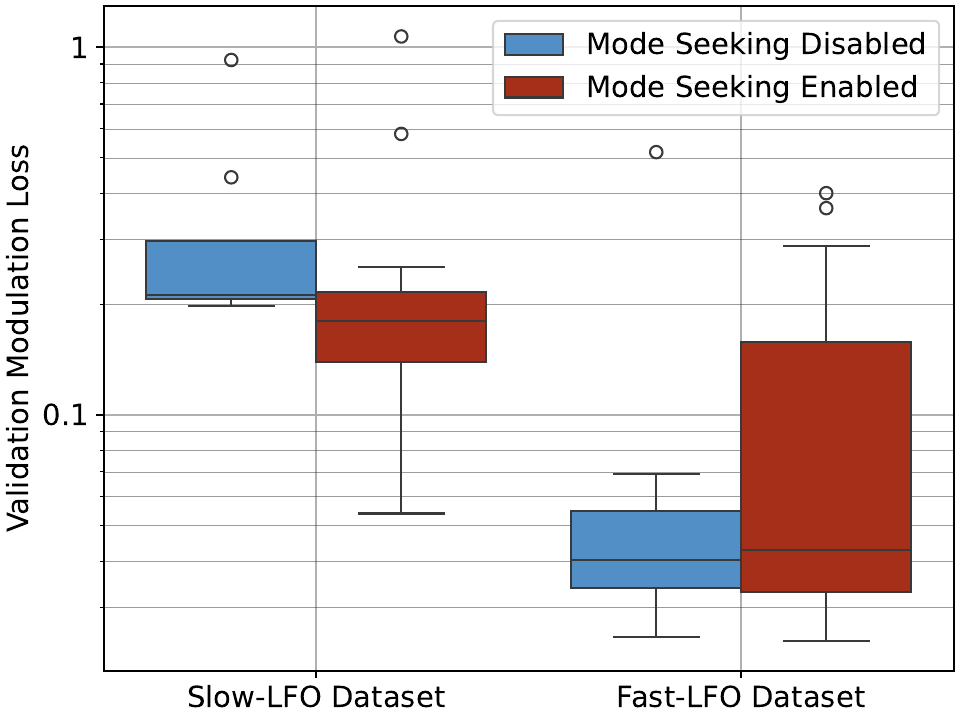}
    \caption{Minimum modulation metric over the validation subset obtained after the adversarial phase (log-scaled axis).}
    \label{fig:g11_modspec}
\end{figure}

\begin{figure*}
    \centering
    \begin{minipage}[c]{0.03\textwidth}
        \begin{sideways}
            Slow-LFO Dataset
        \end{sideways}
    \end{minipage}
    \begin{minipage}[c]{0.92\textwidth}
        \begin{subfigure}[b]{0.245\textwidth}
            \includegraphics[width=\textwidth]{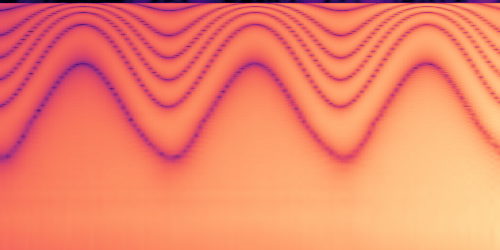}
            \caption{Reference}
            \label{fig:jaes_spec_long_00}
        \end{subfigure}
        \begin{subfigure}[b]{0.245\textwidth}
            \includegraphics[width=\textwidth]{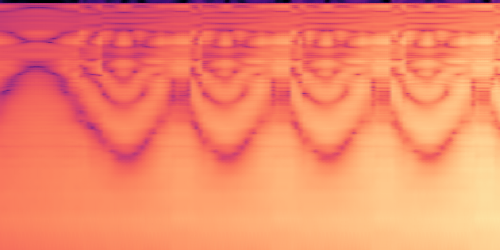}
            \caption{Modulation error: 0.054}
            \label{fig:jaes_spec_long_01}
        \end{subfigure}
        \begin{subfigure}[b]{0.245\textwidth}
            \includegraphics[width=\textwidth]{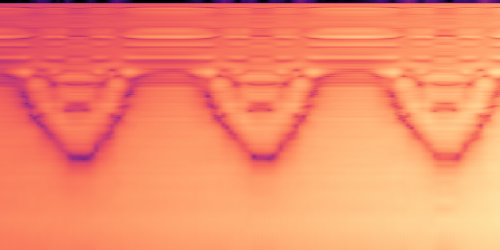}
            \caption{Modulation error: 0.111}
            \label{fig:jaes_spec_long_02}
        \end{subfigure}
        \begin{subfigure}[b]{0.245\textwidth}
            \includegraphics[width=\textwidth]{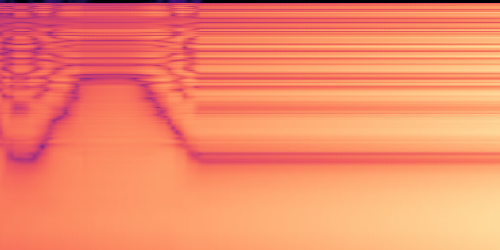}
            \caption{Modulation error: 0.171}
            \label{fig:jaes_spec_long_03}
        \end{subfigure}
        \\
        \begin{subfigure}[b]{0.245\textwidth}
            \includegraphics[width=\textwidth]{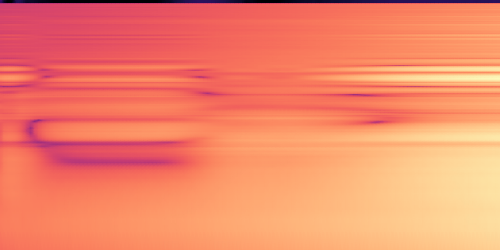}
            \caption{Modulation error: 0.206}
            \label{fig:jaes_spec_long_04}
        \end{subfigure}
        \begin{subfigure}[b]{0.245\textwidth}
            \includegraphics[width=\textwidth]{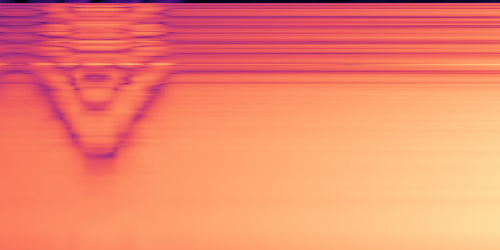}
            \caption{Modulation error: 0.222}
            \label{fig:jaes_spec_long_05}
        \end{subfigure}
        \begin{subfigure}[b]{0.245\textwidth}
            \includegraphics[width=\textwidth]{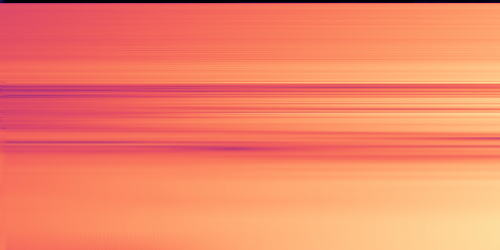}
            \caption{Modulation error: 0.596}
            \label{fig:jaes_spec_long_06}
        \end{subfigure}
        \begin{subfigure}[b]{0.245\textwidth}
            \includegraphics[width=\textwidth]{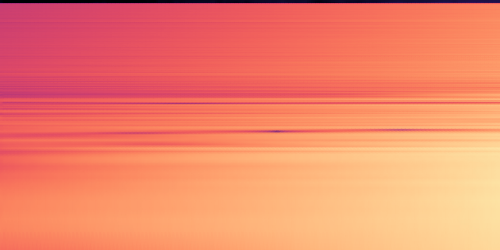}
            \caption{Modulation error: 0.959}
            \label{fig:jaes_spec_long_07}
        \end{subfigure}
    \end{minipage}
    \begin{minipage}[c]{0.04\textwidth}
        \includegraphics[width=\textwidth]{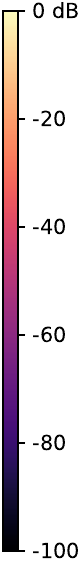}
    \end{minipage}
    \noindent\rule{\textwidth}{1pt}
    \vspace{-0.75em}
    \\
    \begin{minipage}[c]{0.03\textwidth}
        \begin{sideways}
            Fast-LFO Dataset
        \end{sideways}
    \end{minipage}
    \begin{minipage}[c]{0.92\textwidth}
        \begin{subfigure}[b]{0.245\textwidth}
            \includegraphics[width=\textwidth]{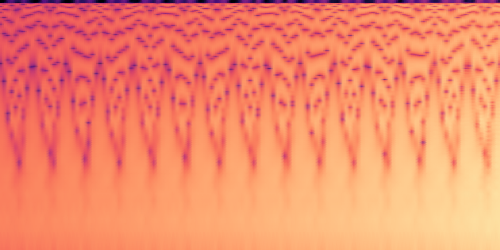}
            \caption{Reference}
            \label{fig:jaes_spec_long_08}
        \end{subfigure}
        \begin{subfigure}[b]{0.245\textwidth}
            \includegraphics[width=\textwidth]{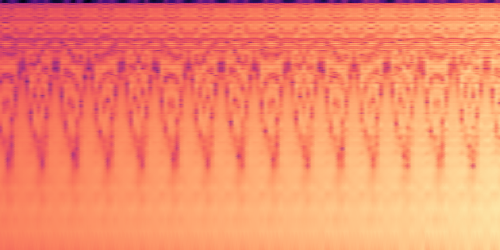}
            \caption{Modulation error: 0.018}
            \label{fig:jaes_spec_long_09}
        \end{subfigure}
        \begin{subfigure}[b]{0.245\textwidth}
            \includegraphics[width=\textwidth]{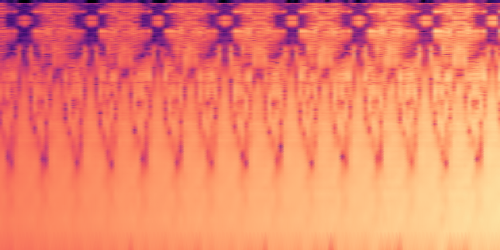}
            \caption{Modulation error: 0.033}
            \label{fig:jaes_spec_long_10}
        \end{subfigure}
        \begin{subfigure}[b]{0.245\textwidth}
            \includegraphics[width=\textwidth]{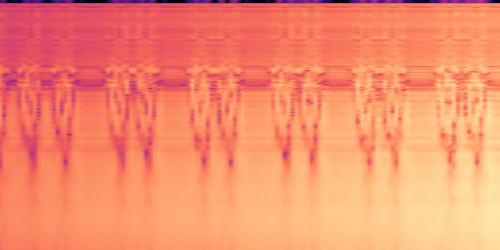}
            \caption{Modulation error: 0.069}
            \label{fig:jaes_spec_long_11}
        \end{subfigure}
        \\
        \begin{subfigure}[b]{0.245\textwidth}
            \includegraphics[width=\textwidth]{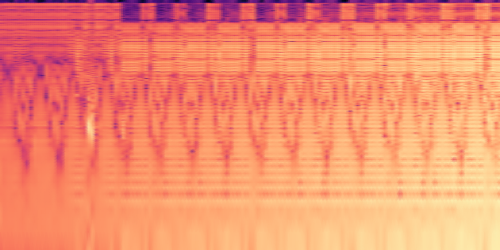}
            \caption{Modulation error: 0.080}
            \label{fig:jaes_spec_long_12}
        \end{subfigure}
        \begin{subfigure}[b]{0.245\textwidth}
            \includegraphics[width=\textwidth]{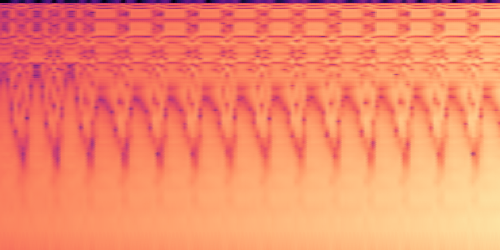}
            \caption{Modulation error: 0.089}
            \label{fig:jaes_spec_long_13}
        \end{subfigure}
        \begin{subfigure}[b]{0.245\textwidth}
            \includegraphics[width=\textwidth]{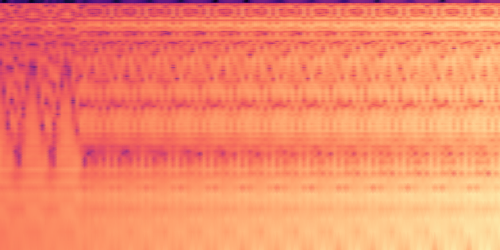}
            \caption{Modulation error: 0.175}
            \label{fig:jaes_spec_long_14}
        \end{subfigure}
        \begin{subfigure}[b]{0.245\textwidth}
            \includegraphics[width=\textwidth]{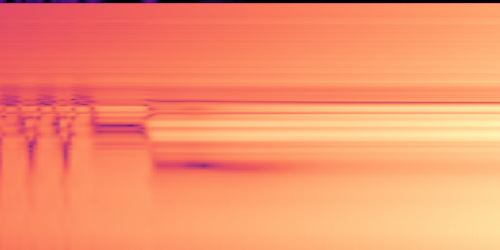}
            \caption{Modulation error: 0.584}
            \label{fig:jaes_spec_long_15}
        \end{subfigure}
    \end{minipage}
    \begin{minipage}[c]{0.04\textwidth}
        \includegraphics[width=\textwidth]{fig/jaes_spec_cbar.pdf}
    \end{minipage}
    \caption{Output spectrograms of cherry-picked models (including best and worst) trained through the adversarial phase on the Slow-LFO (top) and Fast-LFO (bottom) datasets, using the 4-second-long chirp-train input signal, ordered by modulation metric. \specdesc}
    \label{fig:jaes_spec_long}
\end{figure*}

First, each model is trained in the adversarial mode, with the MR-STFT loss weight set to $0.005$. We chose to initialize the states using the angle initialization. We stop training after 400k iterations and select the checkpoint achieving the lowest modulation metric in validation as the resulting model of this phase and starting point for the next one.

Fig. \ref{fig:g11_modspec} depicts the modulation metric reached by all trainings in the adversarial phase.
It is seen that the Slow-LFO phaser effect is harder to train than the Fast-LFO one.
The impact of mode seeking is mixed: it improves modulation metric in the context of a slow LFO, but it seems detrimental for a fast LFO as the values are spread further to the top.
We note that, empirically, values of the modulation metric below 0.2 tend to mean that a periodic modulation close to the one of the target LFO is present, while values above 0.9 tend to represent an absence of modulation. 
Over the Fast-LFO dataset, all the models obtained a modulation metric below $0.4$, indicating they all display a form of modulation.

Fig. \ref{fig:jaes_spec_long} shows the spectrograms of the outputs of some models using the chirp-train signal (and used for the evaluation of the modulation metric). 
They illustrate the potential models one can obtain through adversarial training, and they also highlight the limits of our modulation metric.
This metric is not directly sensitive to the modulation shape, as shown in Fig. \ref{fig:jaes_spec_long_01} and \ref{fig:jaes_spec_long_11} achieving a low modulation error despite the unusual shape of the resulting modulation.
The metric is sensitive to the duration for which the modulation is sustained. While this may appear desirable, it can result in favoring long-duration, unexpected modulation patterns (Fig. \ref{fig:jaes_spec_long_04}) over preferable but non-persisting shapes (Fig. \ref{fig:jaes_spec_long_05}).
Lastly, as can be expected, this metric does not evaluate audio quality or spectral accuracy, and thus a number of undesirable effects are observed such as static attenuations at certain frequencies (as in Fig. \ref{fig:jaes_spec_long_09} and \ref{fig:jaes_spec_long_12}), artefacts (Fig. \ref{fig:jaes_spec_long_10} and \ref{fig:jaes_spec_long_12}), and poor high-frequency processing globally.
Consequently, continuing training using a spectral loss should improve modeling quality.

\subsubsection{Fine-Tuning Phase}
\label{finetuning}

\begin{figure*}
    \centering
    \begin{minipage}[c]{0.03\textwidth}
        \begin{sideways}
            \hspace{3cm}
            Slow-LFO
            \hspace{2em}
        \end{sideways}
        \begin{sideways}
            Fast-LFO
        \end{sideways}
    \end{minipage}
    \begin{minipage}[c]{0.91\textwidth}
        \begin{subfigure}[b]{0.195\textwidth}
            \includegraphics[width=\textwidth]{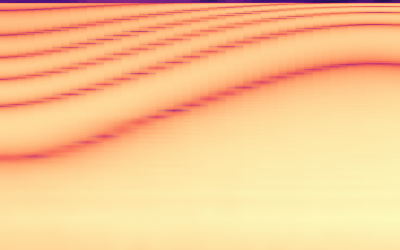}
            \caption{Reference}
            \label{fig:jaes_spec_00}
        \end{subfigure}
        \begin{subfigure}[b]{0.195\textwidth}
            \includegraphics[width=\textwidth]{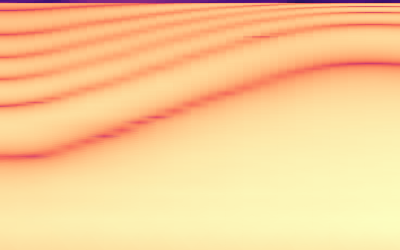}
            \caption{MR-STFT: 0.562}
            \label{fig:jaes_spec_01}
        \end{subfigure}
        \begin{subfigure}[b]{0.195\textwidth}
            \includegraphics[width=\textwidth]{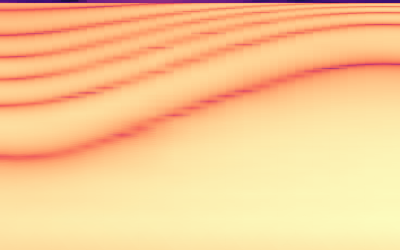}
            \caption{MR-STFT: 0.694}
            \label{fig:jaes_spec_11}
        \end{subfigure}
        \begin{subfigure}[b]{0.195\textwidth}
            \includegraphics[width=\textwidth]{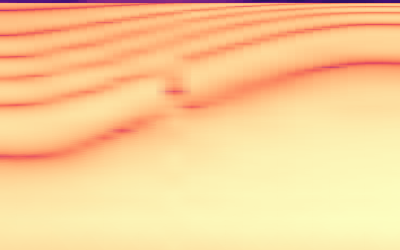}
            \caption{MR-STFT: 0.696}
            \label{fig:jaes_spec_12}
        \end{subfigure}
        \begin{subfigure}[b]{0.195\textwidth}
            \includegraphics[width=\textwidth]{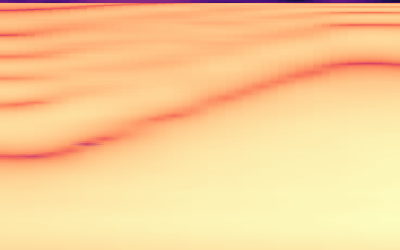}
            \caption{MR-STFT: 0.728}
            \label{fig:jaes_spec_02}
        \end{subfigure}
        \\
        \begin{subfigure}[b]{0.195\textwidth}
            \includegraphics[width=\textwidth]{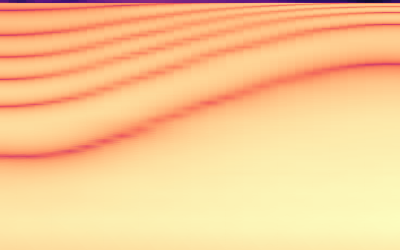}
            \caption{MR-STFT: 0.739}
            \label{fig:jaes_spec_03}
        \end{subfigure}
        \begin{subfigure}[b]{0.195\textwidth}
            \includegraphics[width=\textwidth]{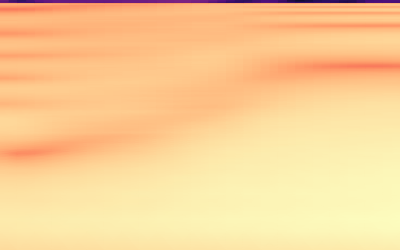}
            \caption{MR-STFT: 0.825}
            \label{fig:jaes_spec_04}
        \end{subfigure}
        \begin{subfigure}[b]{0.195\textwidth}
            \includegraphics[width=\textwidth]{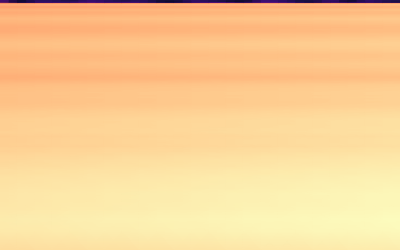}
            \caption{MR-STFT: 0.915}
            \label{fig:jaes_spec_06}
        \end{subfigure}
        \begin{subfigure}[b]{0.195\textwidth}
            \includegraphics[width=\textwidth]{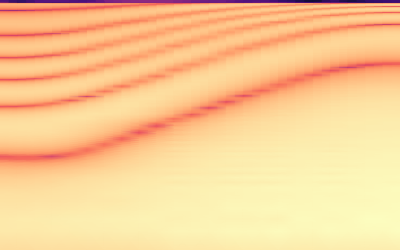}
            \caption{MR-STFT: 0.923}
            \label{fig:jaes_spec_08}
        \end{subfigure}
        \begin{subfigure}[b]{0.195\textwidth}
            \includegraphics[width=\textwidth]{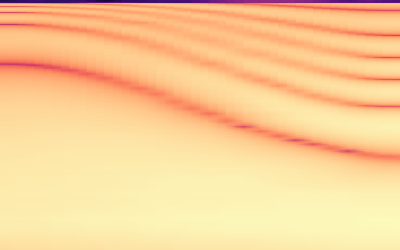}
            \caption{MR-STFT: 1.259}
            \label{fig:jaes_spec_10}
        \end{subfigure}
        \noindent\rule[10pt]{\textwidth}{1pt}
        \vspace{-20pt}
        \\
        \begin{subfigure}[b]{0.195\textwidth}
            \includegraphics[width=\textwidth]{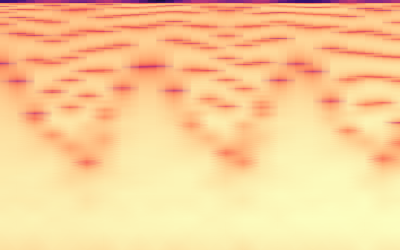}
            \caption{Reference}
            \label{fig:jaes_spec_13}
        \end{subfigure}
        \begin{subfigure}[b]{0.195\textwidth}
            \includegraphics[width=\textwidth]{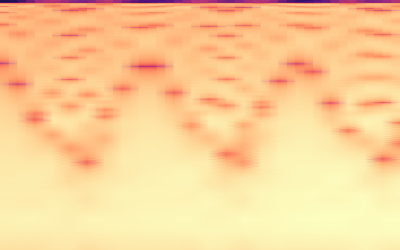}
            \caption{MR-STFT: 0.565}
            \label{fig:jaes_spec_14}
        \end{subfigure}
        \begin{subfigure}[b]{0.195\textwidth}
            \includegraphics[width=\textwidth]{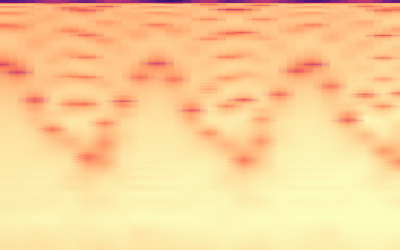}
            \caption{MR-STFT: 0.935}
            \label{fig:jaes_spec_15}
        \end{subfigure}
        \begin{subfigure}[b]{0.195\textwidth}
            \includegraphics[width=\textwidth]{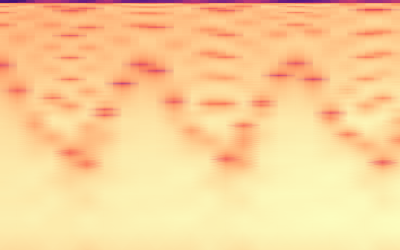}
            \caption{MR-STFT: 0.986}
            \label{fig:jaes_spec_16}
        \end{subfigure}
        \begin{subfigure}[b]{0.195\textwidth}
            \includegraphics[width=\textwidth]{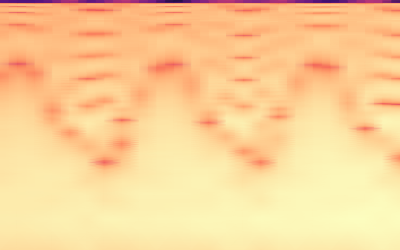}
            \caption{MR-STFT: 1.185}
            \label{fig:jaes_spec_17}
        \end{subfigure}
    \end{minipage}
    \begin{minipage}[c]{0.045\textwidth}
        \includegraphics[width=\textwidth]{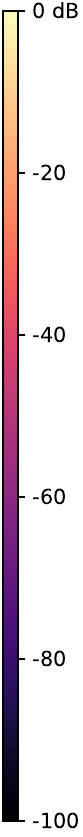}
    \end{minipage}
    \caption{Output spectrograms of selected models during spectral fine-tuning, on the Slow-LFO (top) and Fast-LFO (bottom) datasets, using the chirp-train signal restricted to a single training window (743 ms), ordered by MR-STFT value over this specific window, with (b) and (l) achieving the lowest values among our models. The outputs (c, m) and (d, n) respectively correspond to the models achieving the lowest validation MR-STFT loss, respectively lowest modulation metric. The other outputs were cherry-picked. \specdesc}
    \label{fig:jaes_spec}
\end{figure*}

\begin{figure}
    \centering
    \begin{subfigure}[b]{0.455\linewidth}
        \includegraphics[width=\textwidth]{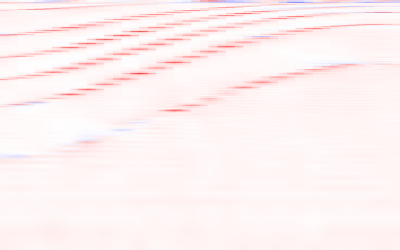}
        \caption{Slow-LFO: \ref{fig:jaes_spec_01} - \ref{fig:jaes_spec_00}}
        \label{fig:diff_slow}
    \end{subfigure}
    \begin{subfigure}[b]{0.455\linewidth}
        \includegraphics[width=\textwidth]{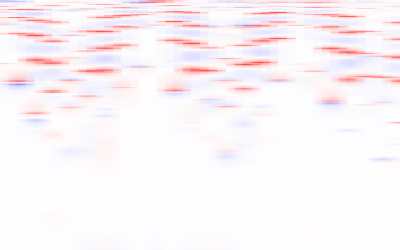}
        \caption{Fast-LFO: \ref{fig:jaes_spec_14} - \ref{fig:jaes_spec_13}}
        \label{fig:diff_fast}
    \end{subfigure}
    \includegraphics[width=0.07\linewidth]{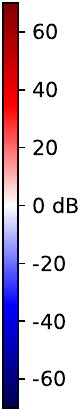}
    \caption{Difference spectrograms comparing the best examples, in terms of the MR-STFT value of the window, against the references for Slow-LFO (a) and Fast-LFO (b).}
    \label{fig:diff}
\end{figure}

\begin{figure}[t]
    \centering
    \includegraphics[width=0.47\textwidth]{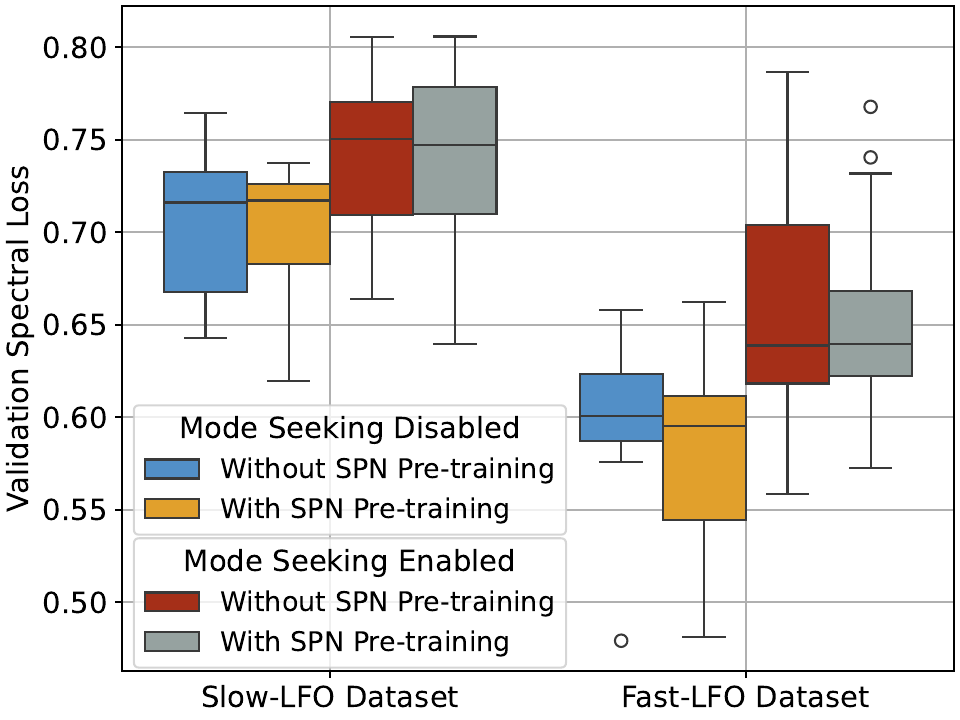}
    \caption{Minimum Validation Spectral (MR-STFT) loss obtained after the fine-tuning phase.}
    \label{fig:mrstft}
\end{figure}

Following the adversarial phase, we proceed to spectral fine-tuning. An SPN is instantiated, and we resume training from the best checkpoint. The discriminator -- except for its feature extraction layers -- and the adversarial losses are disabled, and the FeatBlocks are frozen. The loss function used is the MR-STFT loss, including mode seeking regularization where applicable.

Prior experiments indicated that the sudden addition of a just-initialized SPN could degrade the model. Because the SPN cannot initially predict correct states, the model tends to remove modulation effects entirely, as this is the easiest way to minimize the MR-STFT loss.
Therefore, we investigate the addition of an intermediary phase to pre-train the SPN. 
In this phase, resuming from the adversarial stage, the SPN is introduced while all other model parameters are frozen. The SPN is then trained for 50,000 iterations. At the end of this stage, we select the checkpoint with the lowest validation MR-STFT loss and unfreeze the generator parameters (keeping FeatBlocks frozen). We compare the results with and without this SPN pre-training step, obtained after 400,000 fine-tuning iterations in total.

Fig. \ref{fig:jaes_spec} displays output spectrograms generated using the chirp-train signal, ordered by spectral loss. The models used were taken from the fine-tuning phase, not exclusively from checkpoints that minimized loss. While these results demonstrate significant improvements in quality in many cases, they also highlight the difficulty of training and objectively evaluating time-varying effects.
The quality of the phasing effect appears very similar between Fig. \ref{fig:jaes_spec_01} and \ref{fig:jaes_spec_10}; however, due to a difference in initial phase, the MR-STFT error of the latter is substantially higher than that of the unmodulated signal shown in Fig. \ref{fig:jaes_spec_06}. Even a slight LFO phase shift strongly penalizes the spectral error (Fig. \ref{fig:jaes_spec_08}).
The same observation applies to the Fast-LFO examples; although the modulation quality appears visually decent, the MR-STFT error fluctuates considerably across the outputs in Fig. \ref{fig:jaes_spec_14}, \ref{fig:jaes_spec_15}, \ref{fig:jaes_spec_16} and \ref{fig:jaes_spec_17}.
To better compare the results to the ground truth, we selected the outputs in Fig. \ref{fig:jaes_spec_01} and \ref{fig:jaes_spec_14} which minimize the MR-STFT error, thus favoring a proper alignment, and produced the difference between the output and reference spectrograms, displayed in Fig. \ref{fig:diff}. The notch bands (i.e. narrow attenuation bands characteristic of a phaser effect \cite{kiiskiTimeVariantGrayBoxModeling2016}), especially the higher-order ones, are under-attenuating, both in Fig. \ref{fig:diff_slow} and \ref{fig:diff_fast}. A slight misalignment can also be observed in Fig. \ref{fig:diff_fast}.

The observation of these spectrograms clarifies the objective results presented in Fig. \ref{fig:mrstft}, which depicts the WT MR-STFT loss computed over the validation set for all models and datasets. For each model, the checkpoint retained is the one minimizing the validation MR-STFT subject to a modulation metric below $0.9$. This ensures that the models are compared on realistic checkpoints suitable for our modeling, where modulation occurs, avoiding cases where low MR-STFT errors result from the absence of modulation.
Mode seeking appears detrimental to spectral loss for both datasets, despite the improved modulation metric observed in the adversarial stage for the Slow-LFO dataset. Surprisingly, though we observed more occurrences of mode collapse when mode seeking was disabled in the adversarial stage, we did not find a link between mode collapse and any failure of later fine-tuning. Therefore, this suggests that mode seeking is not necessary when training in multiple phases, and that it should be disabled in the fine-tuning phase. A different formulation of the mode seeking objective or a distinct choice of its hyperparameters might potentially lead to more promising results.
Nonetheless, we observe that the SPN pre-training strategy tends to improve the final spectral loss, especially on the Fast-LFO dataset, and is thus recommended.

As highlighted by Fig. \ref{fig:jaes_spec}, the MR-STFT is highly sensitive to phase and frequency discrepancies in the reproduced modulation, which relies on correct predictions of the SPN. Therefore, the perceptual information given by this metric is mixed with the SPN's generalization capabilities and has limited interpretation when the modulation is not perfectly learned. 
Notably, in preliminary experiments, we noticed that our approach was prone to overfitting, as we regularly observed a decrease in the training MR-STFT loss but not in the validation loss. That is why we introduced an overlap in the division of the dataset into batches (as explained in Section \ref{training_and}), given the short duration of the datasets.

A drawback of our fine-tuning approach is that it only considers minimizing the WT error, i.e. obtaining accurate modulation for the duration of the training window. Therefore, in Streamed Target, the modulation tends to collapse shortly after the duration of a single window, as observed in Fig. \ref{fig:jaes_spec_long_03}, \ref{fig:jaes_spec_long_05}, \ref{fig:jaes_spec_long_14} and \ref{fig:jaes_spec_long_15}. This occurs regularly when modeling the Slow-LFO effect, given that its LFO period is greater than the window duration. However, this is rare for the Fast-LFO dataset, presenting about two LFO periods in a training window. More examples are available on the accompanying website.

\section{CONCLUSION}
\label{conclusion}

In this paper, we presented a novel framework for black-box modeling of time-varying audio effects, addressing the challenge of training without known or aligned modulation signals.
We proposed the SPTVMod architecture, a convolutional-recurrent generative adversarial network, and introduced a promising two-phase training strategy. This begins with an adversarial phase which encourages the model to learn the distribution of the modulation behavior without strict phase constraints; followed by a supervised fine-tuning phase where an SPN adapts the model's internal state to synchronize its modulation with the target data.
To support this process, we introduced a specialized modulation metric based on chirp-train signals and optimal transport. This metric enables the assessment of LFO frequency and presence, while also serving as a stopping criterion for the adversarial phase.
Our experiments on a hardware phaser effect demonstrated that this approach yields models capable of reproducing time-varying behaviors, while black-box supervised methods typically require the knowledge or extraction of modulation signals.

Future work will focus on several key areas.
Originally, in the context of dynamic range compression \cite{bourdinTacklingLongRangeDependencies2024}, the SPN's tendency to overfit was a desirable training mechanism that did not impact inference, which could rely on states initialized with zeros. However, for time-varying effects, reliable state prediction becomes essential for model evaluation, requiring the SPN to present generalization capabilities on unseen data. 
Investigating the underlying structure of these learned states could allow for manual initialization during inference, thereby removing the dependency on the SPN for evaluation and inference.
Furthermore, to address the issue of modulation collapse observed in streamed processing, future architectures could enforce LFO periodicity by employing differentiable oscillators instead of LSTM layers. Alternatively, analyzing the dynamics of the LSTM layers could guide us towards manual control over the modulation.
Beyond LFO-driven units, this framework should be evaluated on the modeling of non-periodic time-varying effects, such as magnetic tape recorders which may present stochastic variations.
A key priority is to investigate the conditioning of SPTVMod by user controls, and to integrate streamed target enhancements as proposed in \cite{bourdinEmpiricalResultsAdjusting2025}, especially to enable the training of slowly evolving modulation effects.
Concurrently, training stability could be improved by investigating new discriminator and SPN architectures.
Finally, the validation of these models requires more rigorous assessment. Our modulation metric could be enhanced by incorporating psychoacoustic models of modulation frequency perception, similar to the modulation spectrum distance proposed in \cite{ramirezGeneralpurposeDeepLearning2019}, and complemented by formal perceptual listening tests to correlate objective loss values with subjective audio quality.

\section{ACKNOWLEDGMENT}

This work is part of a Cifre PhD project funded by ANRT. It benefited from access to the computing resources of the ‘CALI 3’ cluster, operated by the University of Limoges and part of the HPC network in the Nouvelle Aquitaine region, financed by the State and the Region.

\end{document}